\documentclass{PoS}

\title{Rapid and correlated variability of blazar S5 0716+71 from radio- to sub-mm bands}

\ShortTitle{}

\author{\speaker{Lars Fuhrmann}\\
        Max-Planck-Institut f\"ur Radioastronomie, Auf dem H\"ugel 69,
        53121 Bonn, Germany\\
        E-mail: \email{lfuhrmann@mpifr-bonn.mpg.de}}

\author{T. P. Krichbaum, A. Witzel, I. Agudo, S. Britzen, S. Bernhart, V. Impellizzeri, 
        A. Kraus, E. Angelakis, U. Bach, K. Gab\'anyi and J. A. Zensus \\
        Max-Planck-Institut f\"ur Radioastronomie, Bonn, Germany \\
}
\author{S. J. Wagner, L. Ostorero \\
         Landessternwarte Heidelberg-K\"onigsstuhl, Heidelberg, Germany\\ 
}

\abstract{The BL Lac object S5 0716+71 was target of a coordinated and global
multi-frequency campaign to search for rapid and correlated variability and
signatures of the inverse-Compton catastrophe. Here we present first results
obtained from a combined analysis of the cm- to sub-mm observations over a
period of seven days aiming at a detailed study of the intra- to inter-day
variability characteristics and to obtain constraints on the variability
brightness temperatures and Doppler factors comparing the radio data with
the high energy emission recorded by INTEGRAL. A more detailed description 
of the whole cm- to sub-mm observations and our analysis/results will be 
presented in a forthcoming paper. Our analysis reveals the source to be in 
a particular short-term variability phase when compared to the past with a 
correlated $\gtrsim$\,4\,day time scale amplitude increase of up to 35\%, which 
is systematically more pronounced towards higher frequencies. The obtained 
frequency dependent variability amplitudes and time lags contradict expectations 
from interstellar scintillation and strongly suggest a source intrinsic origin 
of this inter-day variability. A 7-day spectral evolution study indicate 
time-variable synchrotron self-absorption and expansion of the emission 
region, consistent with standard models. Assuming relativistic boosting, 
our different estimates of the Doppler factor yield robust lower limits
of $D_{var,IC}$ > 5-22 using the inverse-Compton limit and $D_{var,eq}$ > 8-33
using the equipartition argument. Although high, these values are in good
agreement with Doppler factors obtained from recent VLBI studies and from
the inverse-Compton Doppler factors $D_{IC}$ > 14-16 derived with the X-ray
emission seen by INTEGRAL at 3-200\,keV.}

\FullConference{The 8th European VLBI Network Symposium on New Developments in VLBI Science and Technology
                and EVN Users Meeting  \\
                September 26-29 2006\\
                Torun, Poland}

\begin{document}

\section{Introduction}
\vspace{-0.5cm}
Blazars showing the `classical' IntraDay Variability (IDV) of type II 
in the radio regime (\cite{17}, \cite{6}) display flux 
density variations on time scales of $\lesssim$\,0.5\,--2\,days and variability 
amplitudes ranging from a few up to 30 percent. In order to explain the observed 
variations, intrinsic (e.g. shock-in-jet) as well as extrinsic models 
(i.e. interstellar scintillation) are discussed. If the observed intraday 
variability is interpreted as source intrinsic, the short variability time 
scales imply ultra-compact emission regions with extremely high photon 
densities, leading to apparent brightness temperatures, largely exceeding 
the inverse Compton (IC) limit for incoherent synchrotron radiation 
of $10^{12}$\,K (\cite{7}, \cite{13}) by several orders of magnitude 
(see also \cite{16}). In order to bring these apparent brightness 
temperatures down to the IC-limit, rather extreme relativistic 
Doppler-boosting corrections (with Doppler-factors $\delta > 100$) would be required. 
As such high Doppler-factors (and related bulk Lorentz factors) are not observed in 
AGN jets, alternate jet (and shock-in-jet) models are discussed, which make use
of special, non-spherical geometries and allow for additional relativistic 
corrections (e.g. \cite{15}, \cite{11}). It is also possible 
that the brightness temperatures are intrinsically high and are caused by some 
kind of coherent process (e.g. \cite{3}). Further, it is unclear at 
present, if the IC limit might actually be violated for brief periods of time 
(e.g. \cite{14}). The quasi-periodic and persistent violation 
of this limit (on time scales of hours to days) would then imply subsequent Compton 
catastrophes, each leading to outbursts of inverse Compton scattered radiation via 
second order scattering, which should be observable in the X-/$\gamma$\,-ray bands 
(\cite{7}, \cite{4}). Such efficient cooling flares would then rather quickly 
restore the local brightness temperature in the source (see also \cite{10}). 

In order to search for multi-frequency signatures of such short-term inverse 
Compton flashes occurring at hard X-ray and soft $\gamma$-ray energies,
the proto-typical IDV source S5 0716+71 (z\,>\,0.3) was target of a 
large multi-frequency campaign performed between October 2003 and May 2004.  
The core-campaign was centered around a 500-ksec-INTEGRAL\footnote{INTErnational 
Gamma-ray Astrophysics Laboratory} pointing of the source between November 10 and 
17, 2003. In order to obtain as good as possible time and frequency coverage during 
the INTEGRAL observations, quasi-simultaneous flux density, polarization and VLBI 
observations at radio, millimeter, sub-millimeter, infrared and optical wavelengths 
were organized. An intial description of this large campaign with a first presentation 
of the results of the INTEGRAL observations and selected radio/optical variability curves 
was given by \cite{10}. The analysis of the 3\,mm and 1.3\,mm intensity 
and polarisation measurements performed with the IRAM 30\,m radio-telescope on Pico 
Veleta (PV) were presented by \cite{1}. 

In the following we show and discuss first results of our combined analysis of the radio 
to (sub-) mm total intensity data acquired during the time of the core-campaign 
(November 10--17, 2003). In particular, we concentrate on the intra- to inter-day 
variability behavior of S5\,0716+71 (herafter 0716+714) as deduced from the dense 
in time sampled total intensity data obtained with the Effelsberg 100\,m (EF) and 
the IRAM 30\,m telescope with special regard to a possible violation of the inverse 
Compton limit. These data sets are combined with the available (sub-) millimeter data 
to determine the broad-band variability and spectral characteristics of 0716+714. 
Further, we discuss the possible origin of the observed variations as well as 
present Doppler factors calculated through various methods in comparison with the high 
energy emission observed by INTEGRAL.  
 
A detailed description of all observations and data sets as well as our full analysis and results 
including polarisation will be presented in a forthcoming paper (\cite{5}).     

\section{Observations and data analysis}
\vspace{-0.3cm}
In total, seven radio observatories were involved, covering a frequency range from 
1.4 to 666\,GHz (corresponding to wavelengths ranging from 21\,cm to 
0.45\,mm). In Table \ref{observatories}, the participating observatories 
and their frequency coverage are summarized. Here, we will concentrate on the 
observations performed with EB (4.85, 10.45, 32\,GHz) and PV (86\,GHz). 
At each frequency the target source and a sufficient number of secondary
calibrators were observed continuously with a dense time sampling 
of about two flux density measurements per hour, source and frequency. The data 
reduction was performed following standard procedures for IDV observations with EB 
(e.g. \cite{8}). A detailed description of the PV observations and data reduction 
can be found in \cite{1}. Our individual measurement uncertainties typically lie in
the range of $\le$\,0.5--1\,\% up to 10.45\,GHz and are a factor
2--3 higher at 32\,GHz. At 86\,GHz, a value of $\sim$\,1.2\% was reached
(\cite{1}).    

In order to study the multi-frequency variability characteristics of the source,
each data set was investigated by a statistical variability analysis, based on 
the following steps: (i) a test for the presence of variability ($\chi^{2}$-test), (ii) 
the measurement of the variability amplitudes and (iii) the determination of
the characteristic variability timescales. These methods are described in more
detail by \cite{6}, \cite{12}, \cite{8} and \cite{5}. 
Further, we performed a detailed cross-correlation (CCF) analysis between all 
possible frequency combinations as well as constructed (quasi-) simultaneous
radio spectra by combining all available total intensity data for each 
observing day. Finally, lower limits to the Doppler factor $\delta$ were 
estimated and compared using various methods: (i) $\delta_{var}$ as 
deduced from the variability brightness temperatures, (ii) the equipartition Doppler factor 
$\delta_{eq}$ using calculations of the synchrotron and equipartition magnetic field, 
(iii) $\delta_{var,eq}$ as derived using the equipartition brightness temperature $T_{eq}$
and (iv) the IC Doppler factor $\delta_{IC}$ as calculated from the simultaneous
INTEGRAL observations. Our results will be shortly presented and discussed in the following.  
   
%-----------------------------------------------------------
\begin{table}
\begin{center}
\vspace{-0.5cm}
\caption{Summary of the participating radio observatories and their
  observing wavelengths.} 
\begin{tabular}{l|ll}
\hline
\hline
Radio telescope \& Institute  & Location              & $\lambda_{obs}$ [mm] \\ 
\hline        
WSRT (14x25\,m), ASTRON       & Westerbork, NL        & 210, 180           \\    
Effelsberg (100\,m), MPIfR    & Effelsberg, D         & 60, 28, 9          \\
Pico Veleta (30\,m), IRAM     & Granada, E            & 3.5, 1.3            \\
Mets\"ahovi (14\,m), MRO      & Mets\"ahovi, Finland  & 8                   \\
Kitt Peak (12\,m), ARO        & Kitt Peak, AZ, USA    & 3                   \\
SMTO/HHT  (10\,m), ARO        & Mt. Graham, AZ, USA   & 0.87                \\
JCMT (15\,m), JAC             & Mauna Kea, HI, USA    & 0.85, 0.45           \\
\hline
\hline
\end{tabular}
\label{observatories}
\end{center}
\end{table}
%-------------------------------------------------------------
%______________________________________________________________
%
   \begin{figure}
   \centering
%   \vspace{0.5cm}
   \includegraphics[width=15cm]{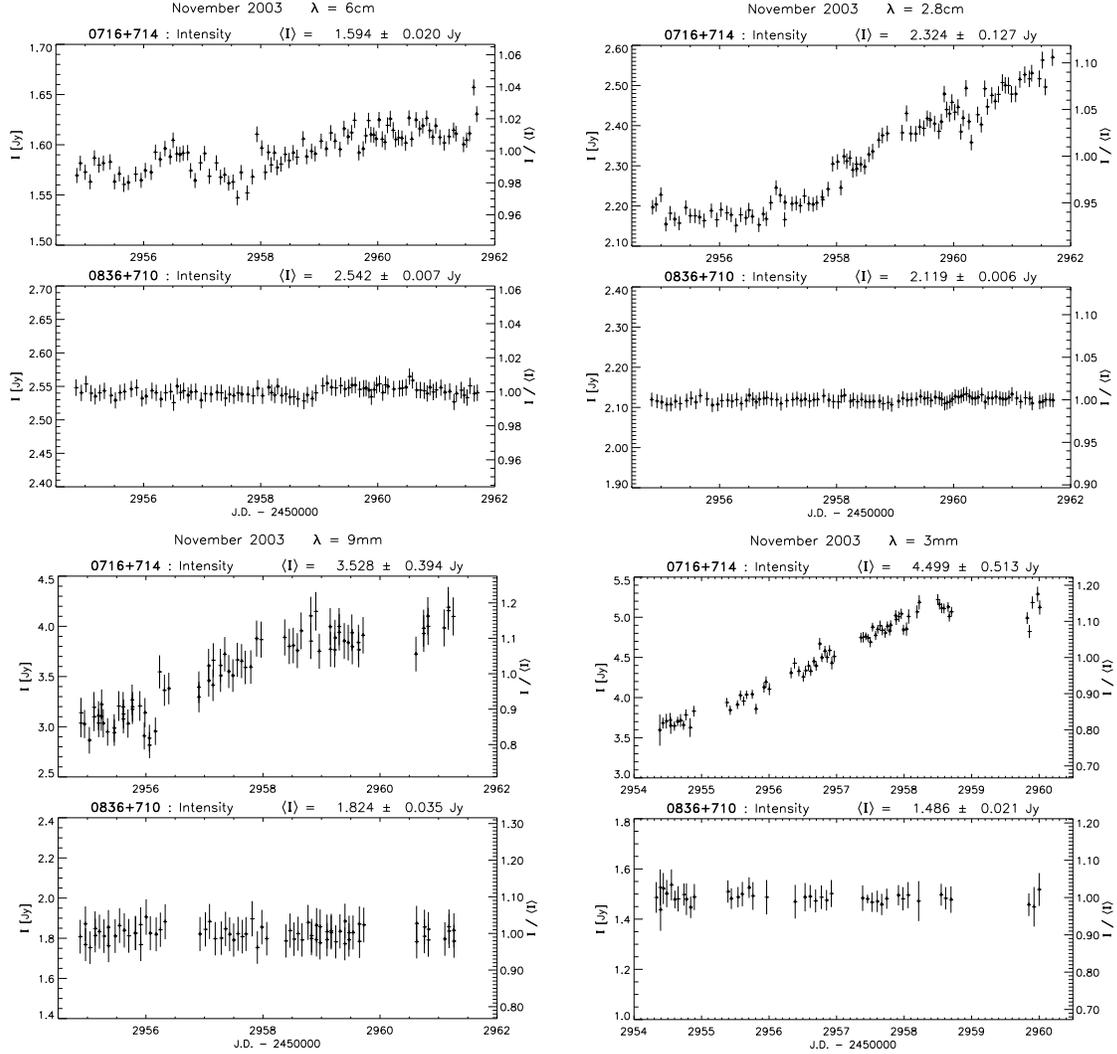}
   \vspace{-0.3cm}
      \caption{Final light curves of 0716+714 obtained at 60, 28, 9 and 3\,mm
        wavelengths, corrected for all systematic effects. For direct comparison, the 
        residual variability of the secondary calibrator 0836+714 is shown at
        each band, which shows remaining peak-to-peak variations 
        of $\sim$\,1\,\% (60, 28\,mm), $\sim$\,8\,\% (9\,mm) and 
        $\sim$\,6\,\% (3\,mm).}
        \label{Fig_LC_all}
   \end{figure}
%______________________________________________________________
%

\section{Results and Discussion}
\begin{figure}
   \centering
   \vspace{-0.5cm}
   \includegraphics[width=15cm,angle=0]{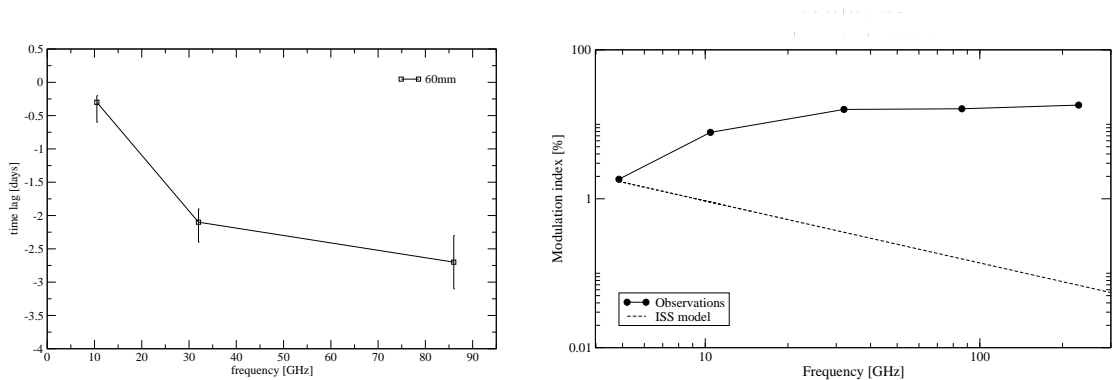}
   \vspace{-0.5cm}
      \caption{Variability behavior of 0716+714 across the observing bands. 
        {\bf Left:} relative time lags for the 60\,mm combinations versus
        frequency (see text). {\bf Right:} observed variability amplitudes 
        versus frequency with model predictions for weak interstellar 
        scintillation superimposed. Note the strong discrepancy between model calculations and observed 
        values.}
         \label{ISS}
   \end{figure}

\vspace{-0.3cm}
In Fig. \ref{Fig_LC_all} we plot the flux density measurements of 0716+714 versus time 
at 4.85, 10.45 and 32\,GHz together with the residual flux density variability
of the secondary calibrator 0836+714. In addition, we add from \cite{1} 
the measurements obtained with the IRAM 30\,m-telescope at 86\,GHz. 
A first inspection of the light curves shows 0716+714 to be strongly variable
when compared to the stationary secondary calibrator. Further, the source
appears to be in a particular short-term variability phase when compared to 
the usually observed 'classical', more rapid IDV pattern. At $\lambda$\,60\,mm 
0716+714 shows only few percent low-amplitude variations followed by a slow 
(2\,day) flux density increase after J.D. 2452957.6. The peak-to-peak
amplitude variations are $\leq 4$\,\%. The light curves obtained at 28, 9 and 
3.5\,mm wavelengths are dominated by a much stronger monotonic increase of the 
flux density over a time range of several days. Here, the $\gtrsim 4$ day time 
scale amplitude increase reaches up to $\sim 35$\,\% at $\lambda$\,3\,mm.
Our CCF analysis confirms that the variability pattern occurs strongly 
correlated across the observing bands with a trend of increasing time lag 
$\tau$ towards larger frequency separation. In Fig. \ref{ISS} (left panel) 
we plot the $\tau_{60/i}$ time lags for the 3 observing bands 
($i=28, 9, 3$\,mm) versus observing frequency. A clear and systematic trend 
towards higher frequencies is seen which demonstrates that the observed 
flux density increase occurs first at the higher frequencies and then 
propagates through the radio spectrum towards lower frequencies. 
In addition, the variability amplitude is systematically more pronounced 
towards higher frequencies as seen from Fig. \ref{Fig_LC_all} as well as 
Fig. \ref{ISS} (right panel).  
Such `canonical' variability behavior provides strong evidence for a source intrinsic 
origin of the observed variability. A similar behaviour is seen in AGN and other compact
radio sources over longer time intervals (weeks to years) and is commonly explained by 
synchrotron-cooling and adiabatic expansion of a flaring component or a shock 
(e.g. \cite{9}). Furthermore, the observed frequency dependence of 
the variability amplitudes contradicts our calculations for a standard model of 
interstellar scintillation which further supports an intrinsic interpretation 
(Fig. 2, right panel). The sudden disappearance of the `classical`, more rapid 
IDV behavior of 0716+714 in the cm-band is most likely caused by opacity effects 
and related to the overall flaring (and high radio-to-optical) state of the 
source during the campaign (see also \cite{10}).
%______________________________________________________________
%
%______________________________________________________________
%
%______________________________________________________________
%   
   \begin{figure}[b]
%   \centering
   \vspace{-0.7cm}
   \includegraphics[width=15cm,angle=0]{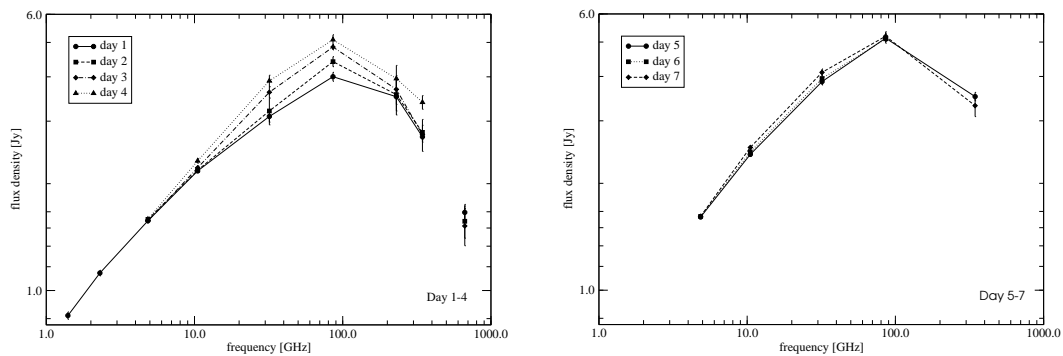}
   \vspace{-0.5cm}
      \caption{Simultaneous cm- to sub-mm spectra of 0716+714 derived from
        daily averages of the flux-density measurements obtained during November 11 to 17 
        using all frequencies. {\bf Left:} the spectra for days 1--4 (November
        11--14, J.D. 2452955-58). {\bf Right:} the spectra of days 5--7 
        (November 15--17, J.D. 2452959-61) are shown below. The flux density 
        errors are small and when invisible, comparable or smaller than the 
        size of the symbols.}
         \label{specs}
   \end{figure}
%______________________________________________________________
%

Our 7-day spectral evolution study shows an inverted synchrotron spectrum peaking near 
$\nu_{m}\sim$\,90\,GHz, with its spectral peak flux continously increasing
during the first 4 days and then remaining nearly constant (see
Fig. \ref{specs}). Slight changes in the daily spectral slopes and the 
turn-over frequency $\nu_{m}$ indicate time-variable synchrotron
self-absorption and expansion of the emission region, consistent with 
standard models. 
The calculated lower limits to the variability brightness temperatures deduced
from the inter-day variations at the different radio-bands exceed the
$10^{12}$\,K inverse-Compton limit by two orders of magnitude in the mm-bands 
and $3-4$ orders of magnitude in the cm-bands. Assuming relativistic boosting, 
our different estimates of the Doppler factor yield robust lower limits of 
$\delta_{var,IC}>\,5-22$ using the inverse-Compton limit and
$\delta_{var,eq}>\,8-33$ using the equipartition argument. Although high, these values 
are in good agreement with Doppler factors obtained from recent VLBI studies 
($\delta_{VLBI}=$\,20--30, \cite{2}) and from the inverse-Compton Doppler 
factors $\delta_{IC}>$\,14--16 we derived using the simultaneous upper limits 
to the source emission in the INTEGRAL 3--200\,keV energy range. The only obtained 
upper limits to the source emission in the soft $\gamma$-ray band (see
\cite{10}) imply high Doppler factors in the source as well as the
non-detection of a simultaneous strong IC catastrophe during the period of our 
observations. Since a strong contribution of interstellar scintillation to the 
observed inter-day variability pattern can be excluded, we thus conclude that 
relativistic Doppler boosting appears to naturally explain the apparent 
violation of the theoretical limits as observed here. 

\small{

}
\end{document}